\documentclass[
    ,final            
  ]
  {aipproc}

\layoutstyle{8x11double}

\begin{document}

\title{Doppler broadening of the iron line and $R-\Gamma$ correlation in
black hole binaries}

\author{M.Gilfanov}{
  address={Max-Planck-Institute f\"ur Astrophysik,
Karl-Schwarzschild-Str. 1, D-85741 Garching, Germany}
,altaddress={Space Research Institute, Profsoyuznaya 84/32, 117997
Moscow, Russia}
}

\author{E.Churazov}{
  address={Max-Planck-Institute f\"ur Astrophysik,
Karl-Schwarzschild-Str. 1, D-85741 Garching, Germany}
,altaddress={Space Research Institute, Profsoyuznaya 84/32, 117997
Moscow, Russia}
}

\author{M.Revnivtsev}{
  address={Max-Planck-Institute f\"ur Astrophysik,
Karl-Schwarzschild-Str. 1, D-85741 Garching, Germany}
,altaddress={Space Research Institute, Profsoyuznaya 84/32, 117997
Moscow, Russia}
}

\begin{abstract}
RXTE/PCA observations of several black hole X-ray binaries in the 
low spectral state revealed a tight correlation between spectral
parameters and characteristic frequencies of variability of X-ray
flux.  In particular, the amplitude of reflection increases and the
slope of Comptonized radiation steepens as the noise frequencies
increase. 
The data also suggest that there is a correlated increase of the width
of the Fe fluorescent line, probably related to the Doppler broadening
of the reflection features. 
Moreover, the width of the line seems to follow $\Delta E/E\propto
\nu_{QPO}^{1/3}$ law. If confirmed with higher energy resolution
observations, this result will have significant impact on the models
of the accretion flow and on our understanding of the nature of
characteristic frequencies of variability in X-ray binaries. 
In particular, it will lend  support to the truncated disk picture,
confirming that the spectral variations are indeed related to changes
of the position of the inner disk boundary and that characteristic
variability frequencies are proportional to the Keplerian
frequency at the inner boundary of the accretion disk.  

\end{abstract}

\maketitle

\begin{figure}
\includegraphics[width=\columnwidth]{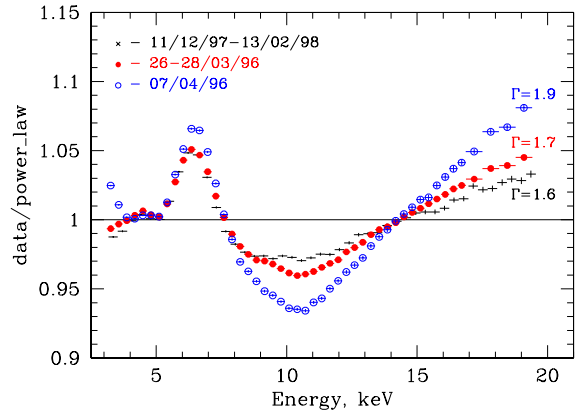}
\caption{ The counts spectra of Cyg X--1 at different epoch 
plotted as a ratio to a power law model. The power law photon index is
indicated against each spectrum.
\label{plrat}
}
\end{figure}

Comptonization of soft seed photons in a hot, optically thin
electron cloud near the compact object is thought to be the main
formation mechanism for hard X--ray emission  in the low
spectral state of accreting black holes \citep{str79,st80}.
Reflection of this Comptonized radiation from neutral or partially
ionized matter of the optically thick accretion disk leads
to appearance of characteristic features in the spectra of X--ray
binaries (Fig.\ref{plrat})  -- the fluorescent
K$_{\alpha}$ line of iron, iron K-edge  and a broad Compton reflection
hump  at higher energies \citep{bst,fab}.  
Their exact shape depends on the ionization state of the matter in the
disk \citep{ionrefl} and might be modified by strong gravity effects
and intrinsic motions in the reflector \citep[e.g.][]{fab1}. Their
amplitude depends on the ionization state and on the solid
angle of the reflector as seen from the source of the primary
radiation. Based on the analysis of a large sample of Seyfert AGNs and
several X--ray binaries \citet{zdz1}, \citep{aaz2003}  found a
correlation between the amplitude of reflection and the slope of the
underlying power law. 

\begin{figure*}
\hbox{
\includegraphics[width=\columnwidth]{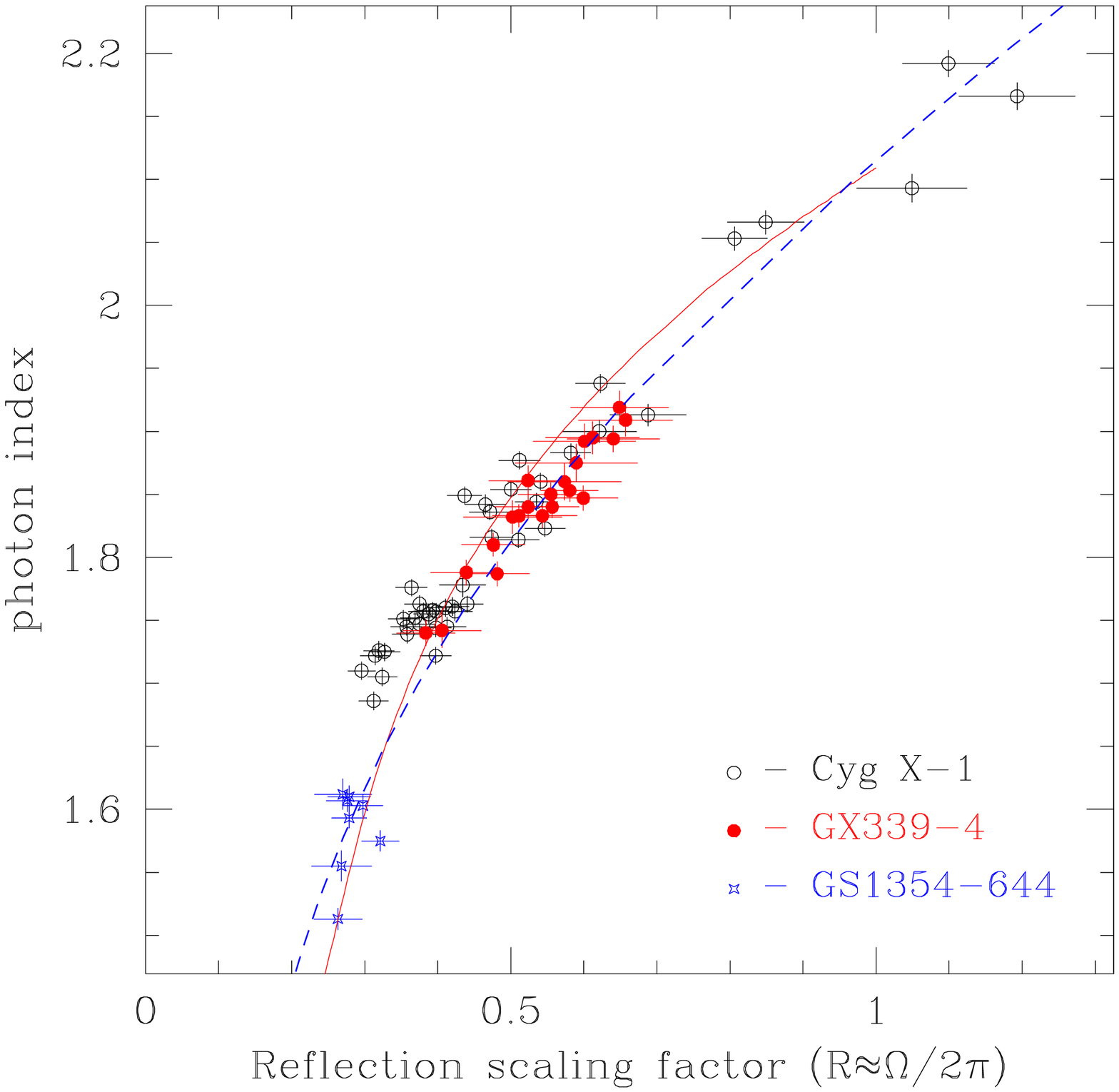}
\includegraphics[width=\columnwidth]{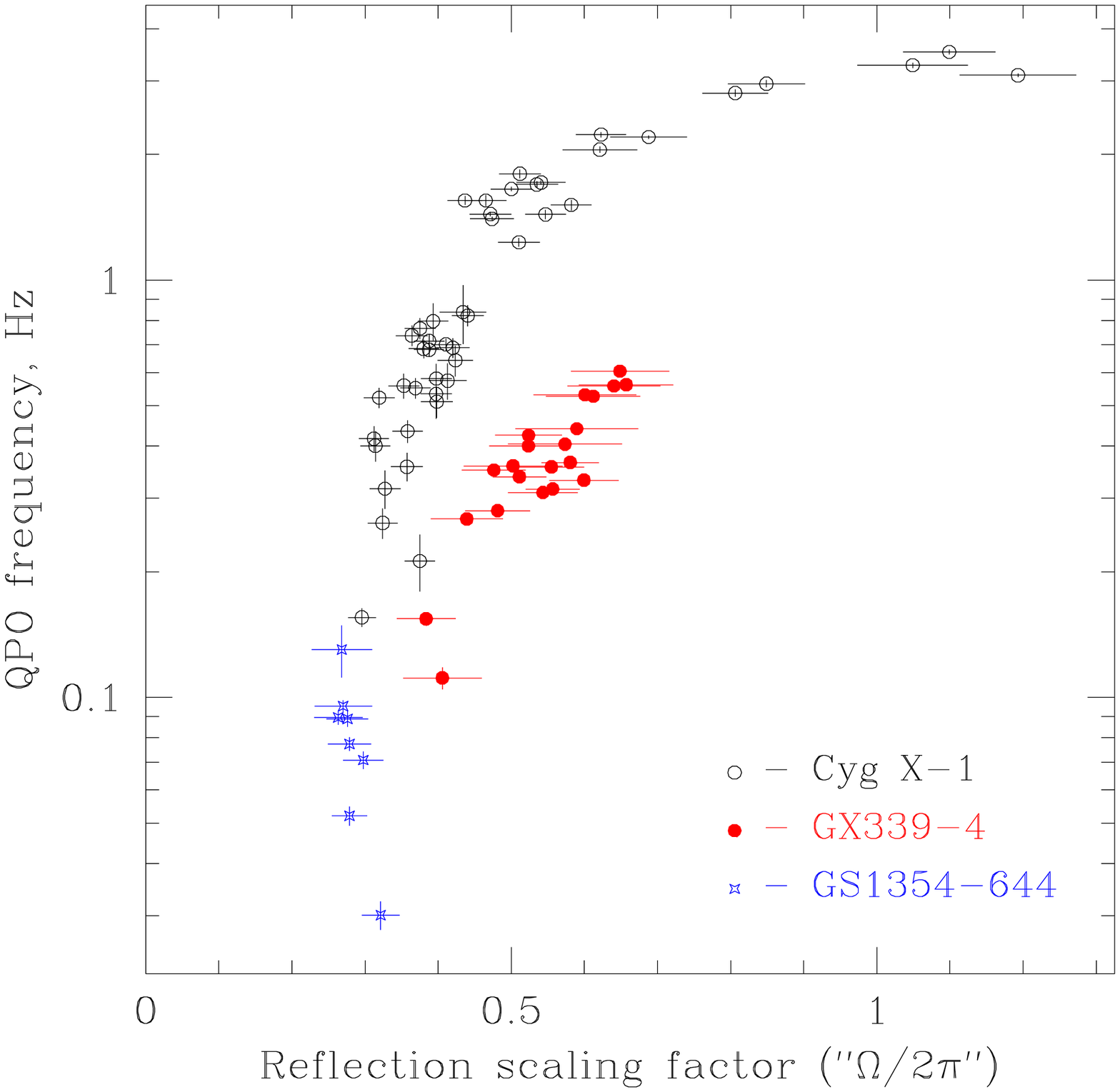}
}
\caption{{\em Left:} Dependence of the photon index of
Comptonized emission $\Gamma$ on the reflection scaling factor $R$. 
The solid and dashed lines show the $\Gamma(R)$ relation predicted in
the disk--spheroid model and in the plasma ejection model
\citep{belob}. See \citet{urumqi} for details of calculations and the
parameters of the models.
{\em Right:} Dependence of the  QPO frequency on the
reflection scaling factor for the same set of observations. 
\label{correlations}
}
\end{figure*}

Power density spectra of black hole binaries in the low state,
plotted in the units of $\nu\times P_{\nu}$, usually appear as a
superposition of two or more rather broad humps containing most of the
power of aperiodic variations below several tens Hz.  These humps
define several characteristic frequencies of variability. Despite of
the fact that the characteristic noise 
frequencies vary from source to source and from epoch to epoch,  they
appear to be  correlated with each other \citep{br_qpo,urumqi}.
Although a number of theoretical models was proposed   to explain the
power spectra of X--ray binaries \cite[e.g.][]{alpar,stella} the
nature of the characteristic noise frequencies is still unclear.

\citet{paper1} and \citet{gx339} showed that a tight correlation
exists between the characteristic noise frequencies  and the spectral
parameters in Cyg X-1 and GX339-4. Based on a large number of RXTE/PCA
observations, they found that the amplitude of reflection increases and
the spectrum 
of primary radiation steepens as the noise frequencies increase.
Interpreting these correlations, \citet{paper1} suggested that increase
of the noise frequency might be caused by the shift of the inner
boundary of the  optically thick accretion disk towards the compact
object. The related increase of the solid angle, subtended by the
disk, and of the influx of the soft photons to the Comptonization
region lead to an increase of the amount of reflection and
steepening of the Comptonized spectrum.

Below we summarize the results on correlation between the spectral
parameters and characteristic frequencies of variability and discuss
the behavior of the Doppler broadening of the reflected 
emission.

\section{Correlation between spectral and timing parameters}

The results presented in Fig.\ref{correlations} are based on the
publicly available data 
of a number of RXTE/PCA \citep{rxte} observations of Cyg X--1, GX339-4
and GS1354--644 performed from 1996--1999. The spectral model and
procedures for spectral approximation and for fitting the power density
spectra are described in \citep{paper1,urumqi,gx339}.
We emphasize, that the  model used for spectral approximation of the
data is obviously oversimplified and does not include a number of
physically important effects. Some of the parameters,
for example, the binary system inclination, were fixed at fiducial
values. Consequently, the best fit values  do not necessarily
represent the exact values of the physically interesting parameters.
Particularly subject to various uncertainties is the reflection
scaling factor $R\sim\Omega/2\pi$, \citep{aaz2003}. 
This might explain the values of $R$
exceeding unity, obtained for some of the spectra. 
However, as was shown in \citep{paper1} and
\citep{gx339}, the model does correctly rank  the spectra according to
the strength of the reflected component and the slope of the
underlying power law.

\section{Toy models}

The photon index $\Gamma$ of the Comptonized radiation is controlled
by the Compton amplification factor $A$ which equals to 
the ratio of the energy deposited into the electrons to the
flux of  soft seed photons to the Comptonization
region. The concrete shape of the $\Gamma(A)$  relation depends on the
ratio $T_{bb}/T_e$ of the temperatures of the seed photons and the
electrons, the Thomson optical depth and the geometry. In the
simplest although not unique scenario  the correlation 
between the amplitude of the reflected component $R$ and the slope of the
Comptonized spectrum $\Gamma$ (Fig.\ref{correlations}) could be understood
assuming that there is a positive correlation between the solid angle
subtended by the reflecting media, $\Omega_{refl}$, and the flux of
the soft photons to the Comptonization region \citep{zdz1}. 
Existence of such correlation is a strong argument in
favor of the reflecting media being the primary source of the soft
seed photons to the Comptonization region. In the absence of strong
beaming effects a correlation between $\Omega_{refl}$ and the seed
photons flux should be expected since an increase of the solid angle
of the disk seen by the hot electrons ($=\Omega_{refl}$) should
generally lead to the increase of the fraction of the disk emission
reaching the  Comptonization region.

\begin{figure}[t]
\includegraphics[width=\columnwidth]{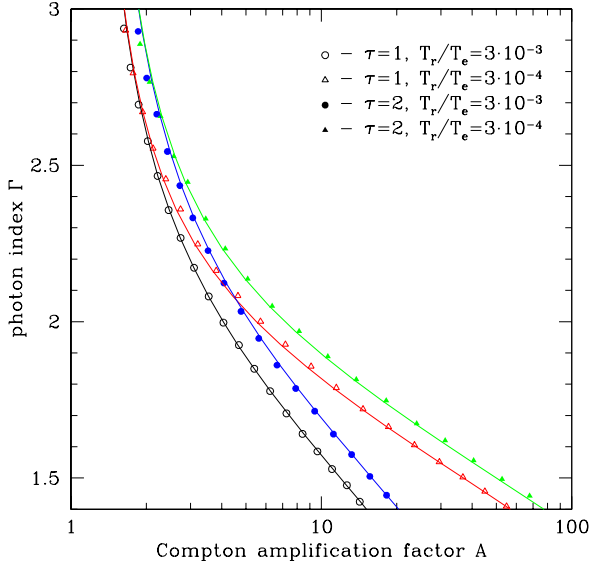}
\caption{Relation between the Compton amplification factor $A$ and
photon index $\Gamma$ of Comptonized radiation. The symbols show  
Monte-Carlo results, the solid curves were calculated according to the
formula given in the text.  
\label{enh_slope}
}
\end{figure}

In order to demonstrate that such geometrical effects can
explain the observed dependence $\Gamma(R)$ we consider two idealized 
models having different mechanism of change of the $\Omega_{refl}$.  
In the first, disk-spheroid model, an optically thin
uniform hot sphere with radius $R_{sph}$ is surrounded with an
optically thick cold disk with an inner radius $R_{disk}$
\citep[e.g.][]{sombrero}, the $\Omega_{refl}$ depending on the ratio
$R_{disk}/R_{sph}$. Propagation of the disk inner edge towards/inwards
the hot  sphere (decrease of $R_{disk}/R_{sph}$) leads to an increase
of the reflection scaling factor, a decrease of the the Compton
amplification factor $A$ and a steepening of the Comptonized spectrum.
In such context the model was first studied by \citet{zdz1} and we
used their results to calculate the relation between the $R$ and
$A$. In the second, plasma ejection model, proposed by \citet{belob},
value of the $\Omega_{refl}$ is defined by bulk motion of the emitting
plasma with mildly relativistic velocity towards or away from the  
disk, which itself remains unchanged. Both models predict relation
between reflection $R$ and $A$ which can be
translated to $\Gamma(R)$  given a dependence $\Gamma(A)$ of the
photon index of Comptonized spectrum on the amplification factor.  
The latter was approximated by:
$$
A=(1-e^{-\tau_T})\cdot\frac{1-\Gamma}{2-\Gamma}\cdot
\frac{\left(\frac{T_e}{T_{bb}}\right)^{2-\Gamma}-1}{\left(\frac{T_e}{T_{bb}}\right)^{1-\Gamma}-1}+e^{-\tau_T}
$$
This formula is based on  a representation of the Comptonized spectrum
by a power law in the energy range $3kT_{bb}-3kT_e$ and agrees with the
results of the Monte-Carlo calculations with reasonable accuracy for
optical depth $\tau_T\sim 1$ and $T_{bb}/T_e\sim 10^{-5}-10^{-3}$
(Fig.\ref{enh_slope}).

\begin{figure*}
\hbox{
\includegraphics[width=\columnwidth]{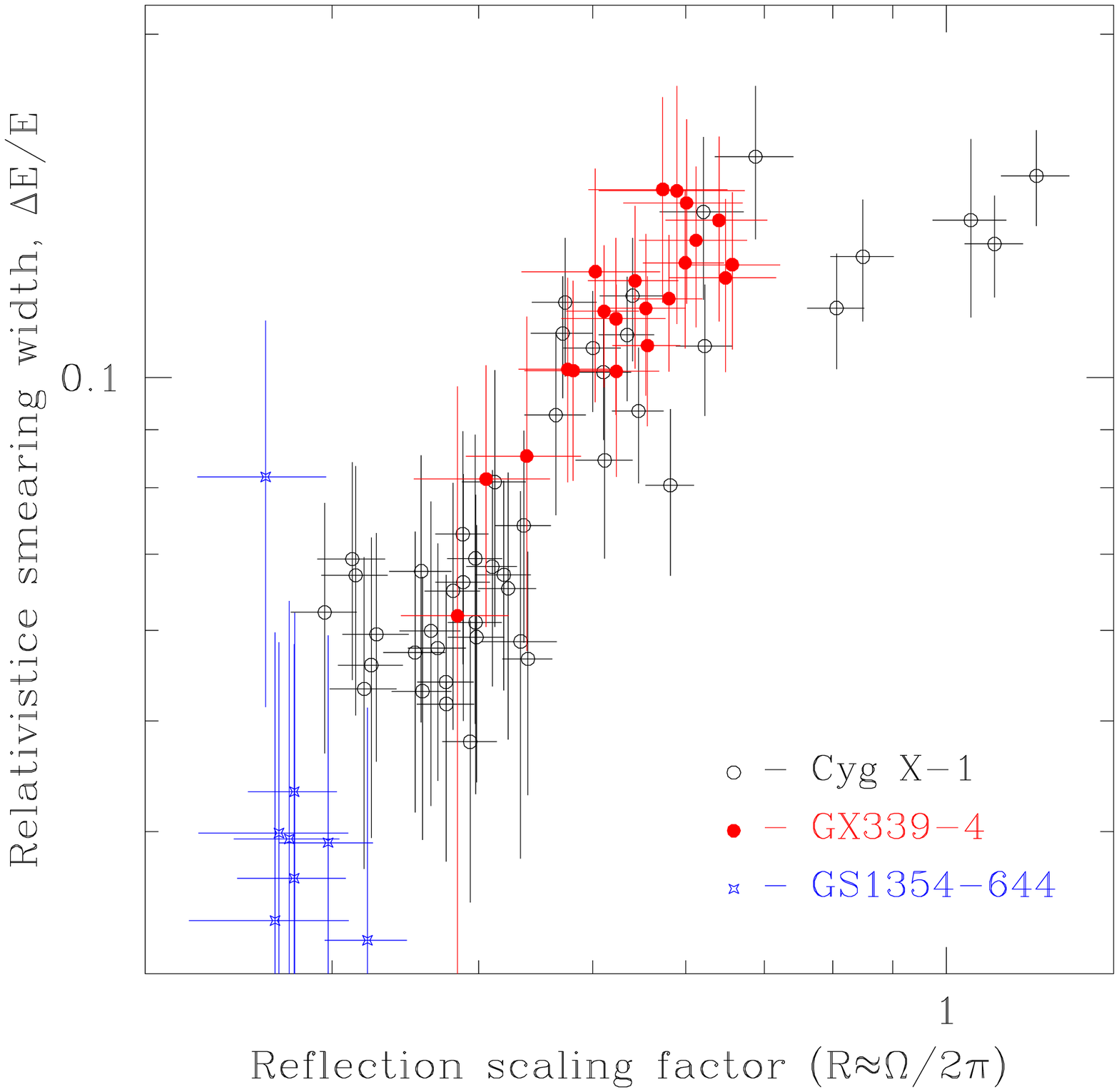}
\includegraphics[width=\columnwidth]{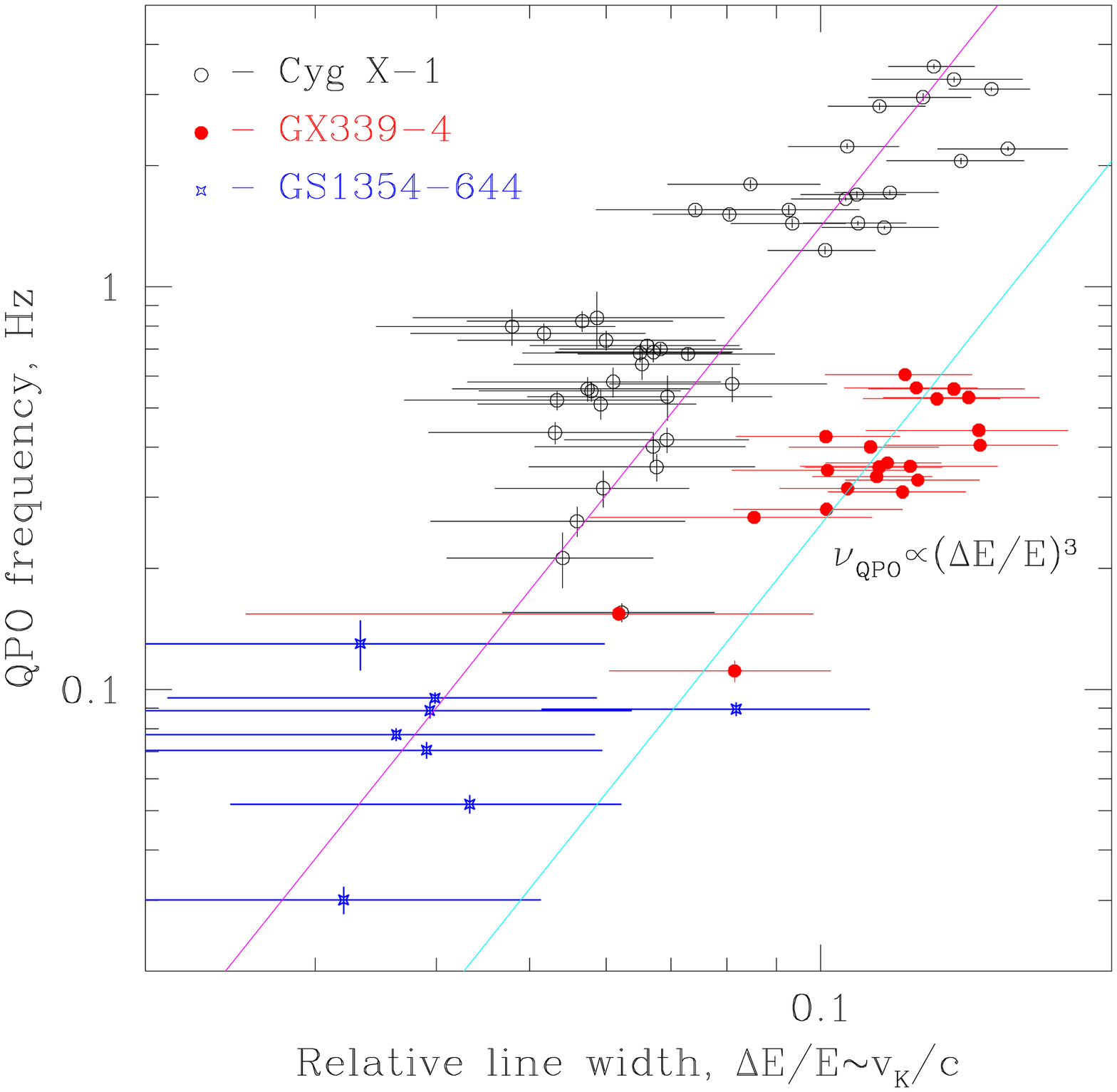}
}
\caption{Relation between the Doppler broadening of the iron
line and the reflection scaling factor ({\em left}) and the
characteristic frequency of variability ({\em right}). 
The straight lines in the right panel show dependence $\Delta
E/E\propto \nu_{QPO}^{1/3}$ expected in the truncated disk picture, if
the characteristic frequencies of variability were proportional to the  
Keplerian frequency at the inner boundary of the accretion disk.
\label{doppler}
}
\end{figure*}

The expected $\Gamma(R)$ relations are shown in 
Fig.\ref{correlations}. With a proper tuning of the parameters both
models can reproduce the observed shape of the $\Gamma(R)$ dependence
and in this respect are virtually indistinguishable. 
The observed range of the reflection $R\sim 0.3-1$ and the slope
$\Gamma\sim 1.5-2.2$ can be explained assuming variation of the disk
radius in from $R_{disk}\sim R_{sph}$ to $R_{disk}\sim 0$ in the
disk-spheroid model or variation of the bulk motion velocity from
$v\sim 0.4 c$ away from the disk to $v\sim 0$ in the plasma ejection model.  
Finally, it should be emphasized that these
idealized models do not include a number of important effects which
might affect the particular shape of the $\Gamma(R)$ dependence.

\section{Doppler broadening of iron line}

The spectrum of emission, reflected from a Keplerian
accretion disk is expected to be modified by the Doppler and special
and general relativity effects \citep{fab1}. If increase of the
reflection is caused by the decrease of the inner radius of the
accretion disk, as predicted in the truncated disk picture,
a correlation should be expected between the amplitude of reflected
emission and its Doppler broadening, in particular the Doppler width
of the fluorescent iron line. Such a correlation is a generic
prediction of  the truncated disk models and might be used to
discriminate between different assumptions about geometry of the
accretion flow.    
The energy resolution of RXTE/PCA is not adequate to 
study relativistic smearing of the reflection features with sufficient 
degree of confidence. However, the data shown in Fig.\ref{doppler}
might present an evidence in favor of a correlated behavior of the
reflection and Doppler broadening of the fluorescent line of iron. 

Speculating further, if the characteristic frequencies of variability
are proportional to the Keplerian frequency at the inner boundary of
the disk, they should scale as 
$$
\nu_{QPO}\propto\omega_K\propto R_{\rm disk}^{-3/2}
$$
As the reflected emission is likely to originate primarily
from the innermost parts of the accretion disk, closest to the source
of Comptonized radiation, the effect of the Doppler
broadening should be proportional to  the Keplerian linear velocity at
the inner edge of the disk:
$$
\frac{\Delta E}{E}\propto \frac{{\rm v}_K}{c}\,\sin i
\propto R_{\rm disk}^{-1/2}\,\sin i
$$ 
Therefore, one might expect, that the characteristic frequencies of
variability and the Doppler broadening of the fluorescent line should
be related via:
$$
\nu_{QPO}\propto \frac{\left(\Delta E/E\right)^3}{M_{\rm BH}\; \sin^3 i}
$$ 
where $M_{\rm BH}$ is the black hole mass and $i$ is inclination of
the binary system. The PCA data indicate, that such dependence
might indeed be the case (Fig.\ref{doppler}). If confirmed with higher
energy resolution observations, this result will have a significant
impact on our understanding of the geometry of the accretion flow in
vicinity of the compact object and of the nature of the characteristic
frequencies of variability. In particular, it will lend  support to
the truncated disk picture, confirming that the spectral variations
are indeed related to changes of the position of the inner disk
boundary and that characteristic variability frequencies are
proportional to the Keplerian frequency at the inner boundary of the
accretion disk.



\bibliographystyle{aipproc}   

\bibliography{sample}

\IfFileExists{\jobname.bbl}{}
 {\typeout{}
  \typeout{******************************************}
  \typeout{** Please run "bibtex \jobname" to optain}
  \typeout{** the bibliography and then re-run LaTeX}
  \typeout{** twice to fix the references!}
  \typeout{******************************************}
  \typeout{}
 }

\end{document}